# Observation of incoherently coupled dark-bright vector solitons in single-mode fibers

**X. Hu,**[1] **J. Guo,**[2] **G. D. Shao,**[2] **Y. F. Song,**[2] **S. W. Yoo,**[1] **B. A. Malomed,**[3] **and D. Y. Tang**[1,*]

[1]*School of Electrical and Electronic Engineering, Nanyang Technological University, Singapore*
[2]*Jiangsu Key Laboratory of Laser Materials and Devices, School of Physics and Electronic Engineering, Jiangsu Normal University, Xuzhou, China*
[3]*Department of Physical Electronics, School of Electrical Engineering, Faculty of Engineering, and Center for Light-Matter Interaction, Tel Aviv University, P.O. Box 39040, Tel Aviv, Israel*

*[*edytang@ntu.edu.sg](mailto:edytang@ntu.edu.sg)*

**Abstract:** We report experimental observation of incoherently coupled dark-bright vector solitons in single-mode fibers. Properties of the vector solitons accord well with those predicted by the respective systems of incoherently coupled nonlinear Schrödinger equations. To our knowledge, this is the first experimental observation of temporal incoherently coupled dark-bright solitons in single mode fibers.

## 1. Introduction

Solitons, i.e. self-trapped nonlinear waves, have been predicted and observed in hydrodynamics [1], plasmas [2], Bose-Einstein condensates [3], biophysics [4], optics [5,6], and in other areas of physics. In nonlinear optics, the soliton formation in single-mode fibers (SMFs) has drawn great interest both due to their significance to fundamental studies and potential applications in long-haul optical communications and optical data-processing systems. The soliton formation in SMFs is described by the nonlinear Schrödinger equation (NLSE), a paradigm model governing the nonlinear-wave propagation in a host of physical systems and giving rise to bright and dark solitons [7,8]. Bright and dark solitons of the NLSE types have been experimentally observed in SMFs [9,10], as well as in water waves [11], atomic Bose-Einstein condensates [12-17], and many other media.

Not only the scalar NLSE but also coupled NLSEs admit soliton solutions. In particular, in the framework of equations modeling the light propagation in birefringent SMFs [18] it has been shown that, due to the cross-phase-modulation (XPM) coupling, bright-bright [19], dark–dark [20], dark-bright (DB) [21-28] vector solitons can be formed. Experimental studies have indeed revealed coupled bright-bright and dark-dark vector solitons [29, 30] in weakly birefringent SMFs. However, to the best of our knowledge, no incoherently coupled dark-bright vector temporal solitons have so far been observed in fibers (Matter-wave DB solitons were experimentally created in binary BECs [15-17]). In this work we report the first experimental evidence of such coupled dark-bright vector solitons in SMFs.

## 2. Theoretical predictions

Under incoherent coupling between two orthogonal linearly polarized modes, the light propagation in a birefringent SMF is governed by the following system of coupled NLSEs:

$$i\frac{\partial u_b}{\partial z} - \beta_b \frac{\partial^2 u_b}{\partial t^2} + \gamma\left(|u_b|^2 + \frac{2}{3}|u_d|^2\right)u_b = 0.$$
$$i\frac{\partial u_d}{\partial z} - \beta_d \frac{\partial^2 u_d}{\partial t^2} + \gamma\left(|u_d|^2 + \frac{2}{3}|u_b|^2\right)u_d = 0. \qquad (1)$$

where $u_b$ and $u_d$ are normalized amplitudes of optical fields carrying the orthogonal linear polarizations, $\beta_b$ and $\beta_d$ are their group-velocity-dispersion (GVD) coefficients, $\gamma$ is the fiber's nonlinearity coefficient, and 2/3 is the standard value of the relative XPM strength. It was first theoretically shown by M. Lisak *et al* [21] that Eq. (1) admits the following coupled DB soliton solutions:

$$u_b = A_b \operatorname{sech}(B\xi) \exp[i(\sigma_b z - \Omega_b t)].$$
$$u_d = A_d \left(1 - C^2 \operatorname{sech}^2 B\xi\right)^{\frac{1}{2}} \exp[i(\sigma_d z - \Omega_d t + \varphi_d)]. \qquad (2)$$

where $\xi = t - \frac{z}{v}$, $\frac{1}{v} = 2\beta_b \Omega_b = 2\beta_d \Omega_d$, $A_b$, $A_d$ and $v$ are arbitrary constants, and

$$B^2 = \frac{1-\alpha^2}{2} \frac{\gamma}{\alpha\beta_d - \beta_b} A_b^2. \qquad (3)$$

$$\sigma_b = \gamma\left[\alpha A_d^2 - \frac{1-\alpha^2}{2} \frac{\beta_b}{\alpha\beta_d - \beta_b} A_b^2\right] + \frac{1}{4\beta_b v^2}. \qquad (4)$$

$$\sigma_d = \gamma\left[\frac{\alpha\beta_b + [(\alpha^2 - 3)/2]\beta_d}{\alpha\beta_b - \beta_d} A_d^2 - \frac{1-\alpha^2}{2} \frac{\beta_d}{\alpha\beta_d - \beta_b} A_b^2\right] + \frac{1}{4\beta_d v^2}. \qquad (5)$$

$$\varphi_d = \frac{N}{2\beta_d A_d^2} \int^\xi \frac{d\xi}{\left(1 - C^2 \operatorname{sech}^2 B\xi\right)^{\frac{1}{2}}}. \qquad (6)$$

$$N^2 = \frac{2\beta_d^2 A_d^6 \gamma}{\beta_d - \alpha\beta_b}(1-\alpha^2)(2-C^2). \qquad (7)$$

$$C^2 = \frac{\alpha\beta_b - \beta_d}{\beta_b - \alpha\beta_d} \frac{A_b^2}{A_d^2}. \qquad (8)$$

Intensity profiles of the coupled dark-bright solitons are plotted in Fig. 1. Different from their scalar counterparts, the coupled dark-bright solitons can be formed in fibers with both normal and anomalous GVD; however, GVD coefficients $\beta_d$ and $\beta_b$ of the dark and bright components should satisfy different conditions in these cases. While in the normal-GVD fiber the condition is $(2/3)\beta_d > \beta_b$, in the case of the anomalous GVD it is $(2/3)\beta_d < \beta_b$. The coupled bright and dark solitons have the same pulse width, which is mainly determined by the strength of the bright component. The darkness of the dark soliton is determined by both the background light intensity and bright-component's strength. If multiple dark-bright solitons are formed in the same fiber, they can have different heights of the bright component and, accordingly, pulse widths.

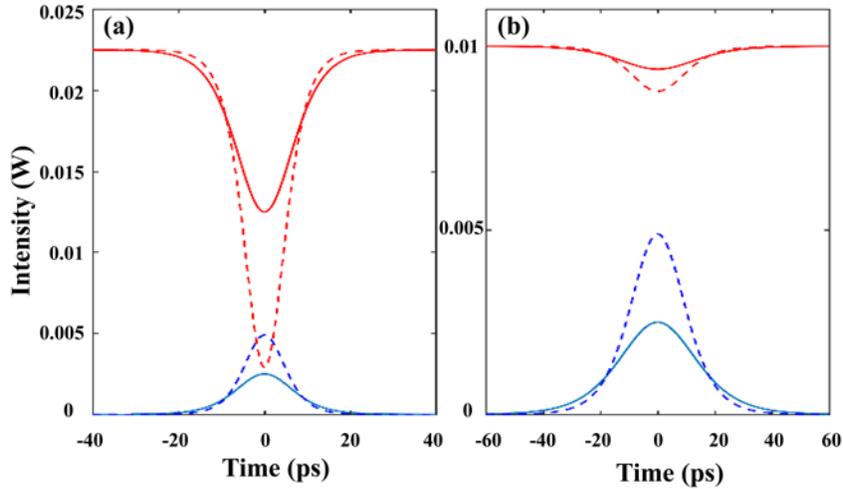

Fig. 1. Typical intensity profiles of the theoretically predicted coupled dark-bright solitons. (a) In the normal-GVD fiber, with $\beta_b = 0.5\text{ps}^2/\text{km}$, $\beta_d = 1\text{ps}^2/\text{km}$, $A_d^2 = 22.5\text{mW}$, $A_b^2 = 2.5\text{mW}$ (solid lines), and $A_b^2 = 4.9\text{mW}$ (dashed lines). (b) In the anomalous-GVD fiber, with $\beta_b = -1\text{ps}^2/\text{km}$, $\beta_d = -0.5\text{ps}^2/\text{km}$, $A_d^2 = 10\text{mW}$, $A_b^2 = 2.5\text{mW}$ (solid lines), and $A_b^2 = 4.9\text{mW}$ (dashed lines).

## 3. Experimental results

Although the coupled dark-bright solitons have been theoretically predicted more than two decades ago [21-27], creation of such complexes in the experiment has not been reported. One of the challenges for the experiment is how to keep the fiber's birefringence constant over a long distance. To overcome the problem, we have adopted a novel approach: instead of letting the light propagate in a long fiber, we circulate it in an active fiber ring cavity. Theoretical studies have shown that, under suitable conditions, the average dynamics of light circulating in the ring is tantamount to that of light propagation in an endless fiber whose parameters are equal to those averaged over the ring [31]. This has an advantage that, through the cavity GVD and birefringence management, one can adjust the effective dispersion and birefringence to any set of required values corresponding to the endless fiber. Indeed, using such a technique, S. T. Cundiff *et al* have demonstrated phase-locked bright vector solitons

[29], and we have demonstrated polarization-rotating vector solitons and higher-order polarization-locked vector solitons in the fibers [32,33].

To realize this approach, we have constructed a fiber ring laser, as schematically shown in Fig. 2. It has a ring cavity consisting of a 3m Er-doped fiber with normal GVD = 64 ps$^2$/km, a 10.5m standard SMF (SMF-28) with anomalous GVD = -24 ps$^2$/km, and a 9m dispersion-compensating fiber with GVD = 5 ps$^2$/km. The fiber laser is pumped by a 1480 nm SMF Raman laser with a maximum pump power ≈ 2W. A polarization-independent isolator is inserted in the cavity to force the unidirectional light circulation. In addition, an intra-cavity polarization controller (PC) is used to fine-tune the linear cavity birefringence. A wavelength division multiplexer (WDM) is used to couple the pump light into the cavity. A 10% fiber coupler is used to output the light.

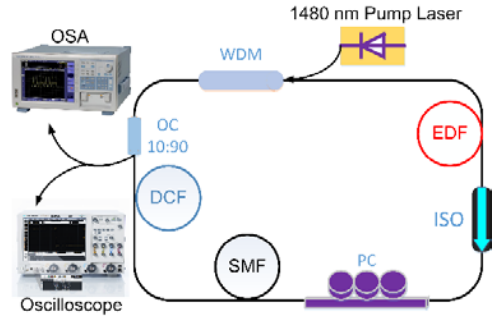

Fig. 2. A schematic of the Er-doped fiber (EDF) ring laser. SMF: Single mode fiber. DCF: Dispersion-compensating fiber. WDM: Wavelength division multiplexer. PC: Polarization controller. ISO: Isolator. OC: Output coupler. OSA: Optical spectrum analyzer.

While building the fiber cavity, care has been taken to ensure that the net cavity birefringence is sufficiently small, so that the laser is simultaneously oscillating in its two orthogonal linearly polarized modes. Incoherent XPM coupling between the two lasing modes can be easily implemented in the experiment. However, limited by the maximum accessible pump power of the laser, we could not create dark-bright solitons when the net cavity GVD was too large. Instead, only the polarization domains and domain-wall solitons [34] could be observed [35]. Therefore, we have switched the laser to operate at a relatively small net averaged cavity GVD, $\left|\overline{\beta_2}\right| < 1.5 \text{ps}^2/\text{km}$.

With the small average cavity GVD, coupled dark-bright solitons could be easily produced in the laser. Fig. 3 shows a typical case of the laser emission, measured at the average cavity GVD $\overline{\beta_2} = -0.67 \text{ps}^2/\text{km}$. Fig. 3(a) displays the polarization-resolved laser emission; Fig. 3(b) presents the corresponding polarization-resolved optical spectra. It is concluded that multiple coupled dark-bright solitons are always initially formed and randomly distributed in the cavity. However, different from the multiple scalar bright solitons, or coupled bright-bright solitons produced in mode-locked fiber lasers [36,37], which always display the same soliton pulse height and energy, different pairs of the DB solitons feature different heights of the bright component and different widths, as seen in the figure. We note that, based on the measured optical spectra of the dark and bright components, they obviously form an XPM-coupled complex.

The net birefringence of the cavity can be slightly changed by tuning paddles of the intracavity PC. At a fixed pump power, carefully increasing the net cavity birefringence, the number of the dark-bright solitons can be significantly reduced, making it possible to produce a single complex of the coupled dark-bright solitons. The complex circulates in the cavity for a long time, until the environmental perturbations essentially alter it, suggesting that the coupled dark-bright solitons are a stable state of the laser operation. We also operated the laser with different values of the anomalous net cavity GVD and pump strength. It is found that, the smaller the net GVD is made, the easier is establishing a robust DB vector-soliton operation regime with a smaller height of the bright component and broader width of the created complex. We note that the formation of the DB vector solitons critically depends on the net cavity birefringence strength. When the net cavity birefringence is large, the frequency offset between the interacting lasing modes also becomes large. In this case no DB vector solitons but only the scalar bright solitons can be observed. However, experimentally we could not switch a DB vector soliton state directly to a scalar soliton state by increasing the net cavity birefringence. In a previous paper we reported a phenomenon of soliton-dark pulse pair formation in a birefringent cavity fiber laser under relatively large net cavity dispersion [38], where scalar bright solitons could be formed in one or both of the orthogonally polarized lasing modes. Because under large average cavity dispersion the formed bright solitons have high peak power, through cross polarization coupling it could induce a broad dark pulse in the CW background of the orthogonal polarization mode. Comparing the DB vector solitons observed in the current paper with that result, although both phenomena are caused by the same effect of cross polarization coupling, due to that their initial conditions and cross coupling strengths are different, the final results are also very different.

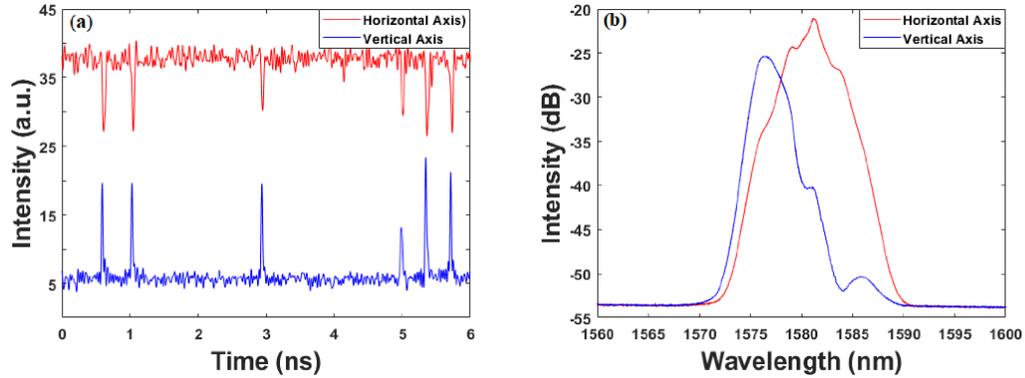

Fig. 3. A typical picture of the emission of XPM-coupled dark-bright solitons, emitted by the fiber laser, in the net anomalous-cavity-GVD regime. a) Oscilloscope traces of the polarization-resolved laser emission; b) The corresponding optical spectra.

To check if the coupled DB solitons could also be formed in the regime of normal net cavity GVD, we operated the laser at $\overline{\beta_2} = 0.67 \text{ps}^2/\text{km}$, selecting this GVD value by cutting away a piece (0.22m) of the SMF. A typical state of the polarization-resolved laser emissions in this case is shown in Fig. 4. Coupled DB solitons with the same features as those seen in Fig. 3 are observed, the measured profiles of the DB solitons hardly showing any obvious difference. The result clearly implies that the formation of the coupled DB solitons is independent on the sign of the net cavity GVD.

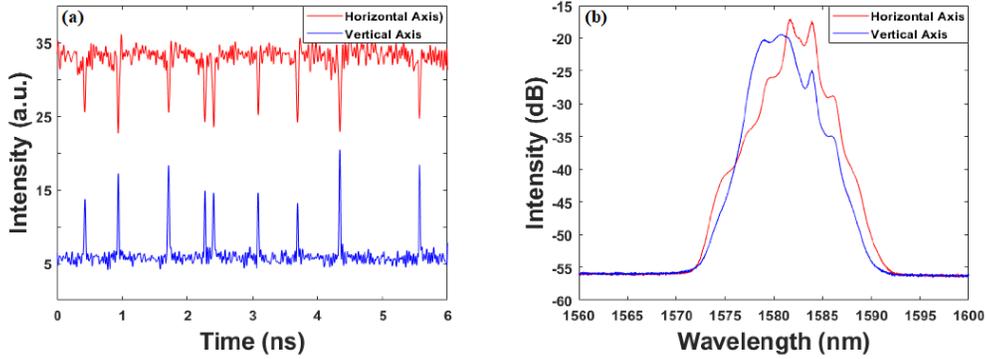

Fig. 4. The same as in Fig. 3, but for the net normal-cavity-GVD regime.

We note that XPM-coupled bright-bright solitons were previously observed in mode-locked fiber lasers [39]. In that case, bright solitons with slightly different group velocities can trap each other through dynamically shifting their central wavelengths. We have found that a similar mechanism applies to the formation of the XPM-coupled dark-bright solitons, in spite of the fact that, without the XPM coupling, a dark (bright) soliton cannot exist in the fiber with anomalous (normal) GVD. It was observed in the experiment that, whenever the dark-bright soliton complexes are formed, a portion of the polarization-resolved spectra in each component dynamically shifts away from their original positions, making the measured spectra in each polarization component asymmetric, as seen in Figs. 3(b) and 4(b). The formation and robust propagation of dark-bright complexes is made possible through this dynamical self-adjustment mechanism. To check the relative intensity of the dark and bright components in each complex, we also measured the total laser output, comparing it in real time with the polarization-resolved laser emissions. It was thus observed that, in the regime with the net anomalous cavity GVD, the total laser emission is a pulse chain identical to the bright-soliton train, while in case of the normal GVD it is a dark-pulse chain identical to the dark-soliton train. This experimental result agrees well with the theoretical prediction, corroborating that the observed complexes are XPM-coupled dark-bright solitons theoretically derived from the XPM-coupled NLSE system.

## 4. Conclusion

In conclusion, we have produced the first experimental observation of XPM-coupled dark-bright vector solitons in SMFs. Features of the observed complexes agree well with those of the dark-bright vector solitons predicted by the XPM-coupled NLSE system. Thus, the results provide the first experimental evidence of the incoherently coupled temporal dark-bright solitons, in accordance with the prediction of the NLSE model. Given that the coupled NLSEs adequately model a variety of physical systems, we expect that the dark-bright vector solitons can be experimentally created in those systems too, under appropriate conditions.

**Funding**